\documentstyle[12pt,epsf]{article}
\begin{document}
\title{Nuclear Response in Electron Scattering at High Momentum Transfer}
\author{Valeriya Gadiyak\thanks{e-mail: Gadiyak@vxinp.inp.nsk.su}\\ and
\\ Vladimir Dmitriev\thanks{e-mail: Dmitriev@inp.nsk.su}\\ Budker
Institute of Nuclear Physics,\\ 630090, Novosibirsk}
\maketitle

\begin{abstract}
The nuclear response functions for high energy electron scattering were 
calculated in the wide region of excitation energy. The three typical 
regions were studied: the quasi-elastic region, the $\Delta$-excitation 
region and the intermediate region where the meson-exchange currents give
significant contribution. In the quasi-elastic and the $\Delta$ regions
the response functions were found for finite size nucleus with account of
relativistic kinematics. The contribution of the meson-exchange currents 
was calculated in the relativistic Fermi-gas model. The results were compared 
with the electron scattering data at high momentum transfer from $^{12}C$ 
and $^{16}O$.
\bigskip

\bigskip

\end{abstract}
\newpage
\section{Introduction}
In a number of experiments on inclusive electron scattering from nuclei,
 $A(e,e')X$, a wide range of the
energy and momentum transfer has been covered
\cite{barr83} - \cite{ang96}.
The inclusive spectrum, as a function of the energy transfer $\omega$ is
characterized by two broad and prominent peaks clearly related to the
processes of quasi-elastic scattering and $\Delta$(1232) resonance
electroproduction.
 Although the other
nucleon resonances are also visible, the $\Delta$ is the most prominent feature
of the transversal nuclear response function and
in the present paper we shall not discuss the region of higher resonances.

In recent years extensive theoretical work has been performed in studies of the 
inclusive electron spectra and the nuclear response functions (see \cite{bof93}
for the full list of references). Although the problem of the Coulomb sum rule 
became less pronounced after the reanalysis of the world data \cite{jou95},
the situation is still not well established and needs further study.
In this connection, the data at the highest available momentum transfer are of
particular interest. At high momentum transfer the role of the nucleon correlations
in nuclear medium decreases and we can hope to reproduce 
the response functions at high momentum transfer within simple independent particle 
models. The fact, that the response functions are  
better reproduced in theoretical calculations at higher momentum transfer
has been already noted, see e.g. \cite{cen97}.

With increasing the momentum transfer, due to mass difference between N
and $\Delta$, the relative distance between two peaks decreases and
they start to overlap. This feature prompts for unified description both
nucleon and $\Delta$ degrees of freedom 
 in this region of excitation energy. At high momentum transfer
the relativistic effects begin to work. This is seen very clearly in
the quasi-elastic region, where the non-relativistic calculations of the
differential cross-section produce a very broad maximum covering not
only quasi-elastic region but the $\Delta$ region as well \cite{alb88}.  We suggest the way,
how the
relativistic kinematics can be accounted for a finite nucleus using formalism of the
response functions. This approach is also used as the base for the
mentioned above unified description of the whole region.

Apart from nucleon and $\Delta$ degrees of freedom the meson-exchange
currents are also necessary to exhaust the cross-section, especially in 
the deep region between $N$ and $\Delta$ \cite{ocon84,kohno81,chlee88}.

In the present paper we develop a formalism that allows to calculate
the response functions of a finite size nucleus with proper account of
the relativistic kinematics for outgoing nucleon or $\Delta$. Full
relativistic treatment of the nuclear response has been done in 
\cite{HorP90}. It was  done for very specific
nucleon- nucleon interaction, within the Walecka model \cite{Wal74}. In
our approach, which is certainly not fully relativistic, we can use all
the variety of the nucleon-nucleus interactions developed for
Schr\"{o}dinger equation. The only correction we make is the
improvement of the relation between the momentum of outgoing particle
and the energy transfer $\omega$. This approach has been already used in the 
calculations of parity violating nuclear response functions \cite{amaro96}.

The paper is organized as follows.  First, we discuss the quasi-elastic
region. We calculate the particle-hole response functions using several
models. Just for the reference we present the calculations for the
non-relativistic, non-interacting Fermi-gas, and non-relativistic
response functions for finite size nuclei. We demonstrate that even for
high $q$ the low-energy wing of the quasi-elastic peak is sensitive to
the finite size effects.

Next, we calculate the response functions for non-interacting, but
relativistic Fermi-gas. It differs by the very important feature: the
differential cross-section in the quasi-elastic region does not broaden in
the relativistic model, and the quasi-elastic peak position is better
reproduced.

Finally, we generalize the calculation of the response functions for
finite nuclei in order to include the relativistic kinematics for
outgoing nucleons. With these response functions we made calculations
of the (e,e') cross-sections for $^{12}$C and $^{16}$O nuclei. We compare 
our calculations with the measured inclusive spectra. This is less informative
than the comparison of the response functions, however, the response functions 
are not yet extracted from the data at this high momentum transfer.
The calculations show that such "quasi-relativistic" model describes the
data at high initial energy somewhat better than the pure relativistic Fermi-gas
model. 

Similar approach has been used for the $\Delta$-excitation. The
$\Delta$-hole response function was calculated for a $\Delta$ moving
in the Woods-Saxon optical potential. The finite size of the target
nucleus is even more important in this case since the kinetic energy of
the $\Delta$ can be small enough providing strong final-state interaction with
the residual nucleus.

Finally, we discuss the deep region and the contribution of the
meson-exchange currents. It was shown \cite{donn75} that the
meson-exchange currents give a significant contribution to the
transversal nuclear response behind the quasi-elastic peak. Extensive 
study of the pionic effects has been performed in \cite{alb90} within the 
static Fermi gas model. We made the calculations in the model of 
relativistic Fermi-gas with account of the nonstatic $w$-dependence
in the spirit of \cite{tjon92}. We found that the meson-exchange
currents (MEC) contribution in the differential cross-sections is
significant. It improves considerably the agreement of the calculations
with the data.

\section{Low energy and quasi-elastic region.}

The inclusive $(e,e')$ cross-section from an unpolarized target in
terms of Coulomb and transversal structure functions is given by
\begin{equation} \label{1.1}
\frac{d^2 \sigma}{d \omega d \Omega}=
\sigma _M \Big [ \frac{q^4}{{\bf q}^4} S_C(\omega,{\bf q})+ \Big (-
\frac{q^2}{2{\bf q}^2}+ \tan ^2(\theta /2) \Big )S_T(\omega,{\bf q})
\Big ],
\end{equation}
where $\sigma _M$ is the Mott cross-section, $\omega =E-E'$ is the
electron energy loss, $\bf q$ is the tree-momentum transfer, $q^2=
\omega ^2-{\bf q}^2$ is the four-momentum transfer squared, and 
$\theta$ is the electron scattering angle.

        The structure functions $S_C(\omega,{\bf q})$ and
$S_T(\omega,{\bf q})$ are related to the imaginary parts of the
corresponding response functions
\begin{equation} \label{1.2}
S_C(\omega,{\bf q})=-
\frac{1}{\pi}\Im m\int_ {-\infty}^ \infty dte^{\imath \omega t}
<T(\rho ^\dagger({\bf q},t) \rho ({\bf q},0))>,
\end{equation}
\begin{equation} \label{1.3}
S_T(\omega,{\bf q})=-
\frac{1}{\pi}
(\delta _{ij}-
\frac{q_iq_j}{{\bf q}^2})\Im m\int_ {-\infty}^ \infty dte^{\imath \omega t}
<T(J^\dagger_i({\bf q},t)J_j({\bf q},0))>,
\end{equation}
where $\rho ({\bf q},t)$ and $J_i({\bf q},t)$ are the Fourie components 
of the charge and electromagnetic current densities.

The Coulomb response function $R_C(\omega,{\bf q})$ for the
non-relativistic Fermi gas is given by
\begin{equation} \label{1.4}
R_C(\omega,{\bf q}) = \sum_{{\bf p},\sigma}\frac{n_{\bf p}- n_{{\bf p}+{\bf q}}}{\omega +
\epsilon_{\bf p} -\epsilon_{{\bf p}+{\bf q}} +\imath\delta},
\end{equation}
where $n_{\bf p}$ is the proton occupation number and $\epsilon_{\bf p}={\bf p}^2/2m$.

For finite size nucleus the response function depends on two momenta
${\bf q}$ and ${\bf q}^\prime$. It can be presented for a spherical
nucleus as
$$
R_C (\omega,{\bf q},{\bf q}') = 16\pi^2\sum_{JM}Y^*_{JM}(\hat{\bf q})
Y_{JM}(\hat{\bf q}')\cdot
$$
\begin{equation}\label{1.5}
 \int_0^\infty r^2drr'^2dr' j_J(qr)j_J(q' r') A_J(\omega;r,r'),
\end{equation}
where the particle-hole propagator $A_J(\omega;r,r')$ is
$$
A_J(\omega;r,r') = \frac{1}{2J+1}\sum_{\nu_1\nu_2}n_{\nu_1}R_{\nu_1}(r)R_{\nu_1}(r')
|\langle l_1j_1||Y_J||l_2j_2\rangle|^2
$$
\begin{equation}\label{1.6}
(G_{l_2j_2}(r,r';\epsilon_{\nu_1}+\omega) + G_{l_2j_2}(r,r';\epsilon_{\nu_1}-\omega)).
\end{equation}
Here $G_{l_2j_2}(r,r';\epsilon)$ is the Green function of the radial
Schr\"{o}dinger equation with the appropriate boundary conditions at
infinity, $R_{\nu}(r)$ is the radial wave function of the occupied
bound state $|\nu\rangle$, and $\langle l_1j_1||Y_J||l_2j_2\rangle$ is
the reduced matrix element of the spherical harmonics
$Y_{JM}(\theta,\phi)$. The structure functions (\ref{1.2},\ref{1.3})
are related to the diagonal part of the response (\ref{1.5}) by
\begin{equation}\label{1.7}
S_C(\omega,q) = -4F^2_e(q^2)\sum_J(2J+1)\int_0^\infty r^2drr'^2dr' j_J(qr)j_J(q r')
\Im m A_J(\omega;r,r'),
\end{equation}
where $F_e(q^2)$ is the proton charge formfactor. Note, that the
imaginary part is nonzero only for the $+\;\omega$ term in (\ref{1.6}).
For the transversal response the similar expressions were obtained.
They differ from (\ref{1.6}) by the tensor operators in the
corresponding particle-hole response functions. The explicit
expressions can be obtained from
\begin{equation}\label{1.8}
{\bf J}({\bf q})= \sum_{{\bf p},\sigma, \sigma^\prime}
F_e(q^2)\frac{2{\bf p}+ {\bf q}}{2m}
a^\dagger_{{\bf p}+ {\bf q},\sigma}a_{\bf p,\sigma}+ F_m(q^2)
\frac{\mu}{2m}[{\bf q} \times a^\dagger_{{\bf p}+ {\bf q},\sigma}
\mbox{\boldmath $\sigma$}_{\sigma \sigma^\prime}a_{{\bf p}, \sigma^\prime}].
\end{equation}

The calculations of the cross-section for free Fermi-gas and the finite
size nucleus are shown in Fig.1. Here we clearly see the
difference in the shape of the quasi-elastic
peak for these two cases. The difference in the shape reflects the difference in the momentum 
distributions of the nucleons in these two models.
The peak is more narrow for the finite size case although the position of the maximum is the
same in both models. Notice, that the sum of all contributions produces the peak slightly wider
in the quasi-elastic region than the data even at this lowest electron energy $E=961$ MeV. For 
higher electron energy the nonrelativistic calculation produces unreasonably wide quasi-elastic
peak. This reflects the necessity of using the relativistic kinematics for these high values 
of the momentum transfer \cite{alb88}.

Let us compare now the response functions for the nonrelativistic and the relativistic Fermi-gas.
For the Coulomb and the transversal responses of the nonrelativistic Fermi-gas
we have
\begin{equation}\label{1.9}
R_C(\omega, {\bf q})=
2 \int
\frac{d^3p}{(2 \pi)^3}
\frac{n_{\bf p}-n_{{\bf p}+ {\bf q}}}
{\omega -{\bf q}^2/2m-
({\bf pq})/m+\imath\delta
\omega / \vert \omega \vert}
\end{equation}
and
$$
R_T(\omega, {\bf q})=\frac{2}{(2 \pi)^3}\int
d^3p
\frac{n_{\bf p}-n_{{\bf p}+ {\bf q}}}
{\omega -
{\bf q}^2/2m-
({\bf pq})/m+\imath \delta
\omega / \vert \omega \vert} \cdot
$$
\begin{equation}\label{1.10}
\Big [
F_e^2 \frac{2}{m^2}({\bf p} -
\frac{{\bf q}({\bf pq})}{{\bf q}^2} \Big )^2+ \frac{F_m^2}{m^2} {\bf q}^2 \Big ].
\end{equation}

\begin{figure}
\epsfxsize 12cm \epsfysize 8cm \epsffile{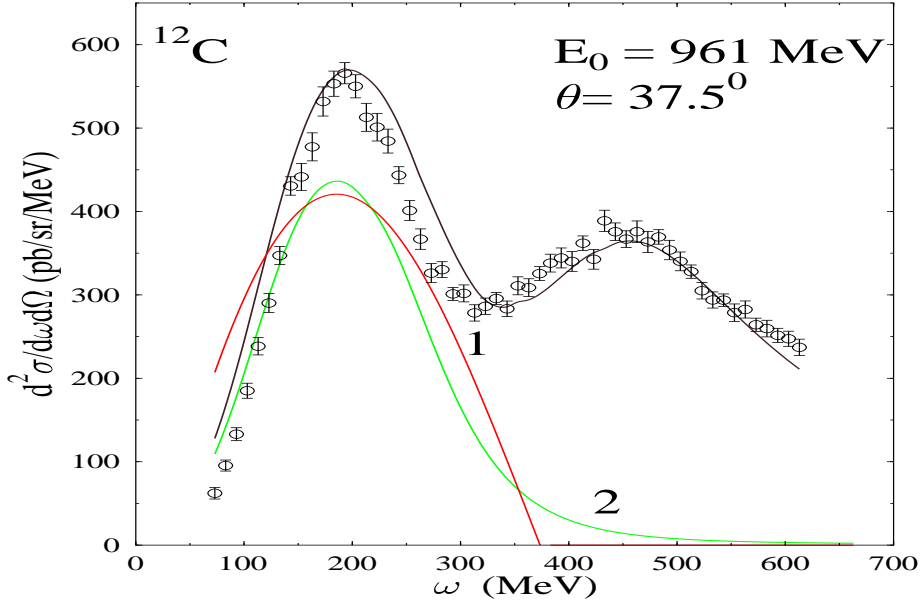}
\caption{The quasi-elastic peak, 1 - free Fermi-gas, 2 - finite size nucleus. The upper line
is the full calculation for nonrelativistic finite size nucleus including quasi-elastic, $\Delta$
, and MEC contributions}
\end{figure}

        In the case of relativistic nucleon we have for the
electromagnetic current density:
$$
J_ \mu (x) = \overline \psi (x)\Gamma_\mu \psi (x),
$$
where
$$
\Gamma_\mu=F_1 \gamma _\mu-
\frac{F_2}{2m} \sigma_{\mu \nu}q^ \nu.
$$
The Dirac formfactors $F_1$ and $F_2$ are related to the charge
and magnetic formfactors by
\begin{equation}\label{1.16}
F_1=
\frac{F_e+ \eta F_m}{1+ \eta}, \quad
F_2=
\frac{F_m-F_e}{1+ \eta}, \quad
\eta = -
\frac{q^2}{4m^2}.
\end{equation}

For the Coulomb and the transversal responses we find then
\begin{equation}\label{1.17}
R_C(\omega, {\bf q}^2)=
\frac{2}{(2 \pi)^3} \int d^3p \Big[
\frac{A({\bf q}^2, \omega)}
{\omega +
q^2/2m- ({\bf pq})/m}+
\frac{A({\bf q}^2,- \omega)}
{-\omega +
q^2/2m+({\bf pq})/m} \Big ]
\end{equation}
\begin{equation}\label{1.18}
R_T(\omega,{\bf q}^2)=
\frac{2}{(2 \pi)^3} \int d^3p \Big[
\frac{A^\prime({\bf q}^2, \omega)}
{\omega +
q^2/2m- ({\bf pq})/m}+
\frac{A^\prime({\bf q}^2,- \omega)}
{-\omega +
q^2/2m+({\bf pq})/m} \Big ],
\end{equation}
where
\begin{equation}\label{1.19}
A({\bf q}^2, \omega)=
\frac{F_e^2}{1+ \eta}
(1 + \frac{\omega}{2m})^2
+F_m^2 \Big [
\frac{\eta}{1+ \eta}
(1+ \frac{\omega}{2m})^2-
\frac{{\bf q}^2}{4m^2} \Big ],
\end{equation}
\begin{equation}\label{1.20}
A^\prime ({\bf q}^2, \omega )=
\frac{F_e^2+\eta F_m^2}{1+ \eta} \cdot
\frac{2}{m^2}  \Big[{\bf p}^2 - \frac {(m\omega 
+q^2/2)^2}{{\bf q}^2} \Big ]
- F_m^2 \frac {q^2}{m^2}.
\end{equation}
The Eqs. (\ref{1.17},\ref{1.18}) were obtained assuming nonrelativistic motion of the
nucleons inside the nucleus, putting thus $E_{\bf p} = m$ whenever it is possible. 
For this reason, we expect the difference between (\ref{1.17}) and its nonrelativistic 
analog (\ref{1.9}) to be mainly in kinematics.
Comparing (\ref{1.9}), (\ref{1.17}) and (\ref{1.10}), (\ref{1.18}) we
find the difference 
in the following. First, in the denominators of (\ref{1.9}) and (\ref{1.10})
we have to substitute $\omega \rightarrow \omega + \omega^2/2m$. Second,
instead of simple formfactors squared we have the factors that are the
combinations of the formfactors with some other factors depending on $\omega$ and
$q^2$. These factors give the difference between Rosenbluth formula for
the electron scattering cross-section from the proton and its nonrelativistic
analog. It is worth to note that with these factors we have now the contribution
to the Coulomb response from neutrons via the magnetic formfactor. At high
momentum transfer this contribution is not negligible.

The substitution of $\omega$ is even more important. Due to this change the
quasi-elastic peak is not broadening and keeps the correct peak position
and the shape compared to the nonrelativistic response. In Fig.2 we see that
the nonrelativistic response produces very broad quasi-elastic peak
in comparison to the relativistic Fermi-gas response. This is not surprising
because now we have the correct relativistic relation between the energy transfer and
the particle-hole energy in the final state. After the substitution we have for the kinetic
energy of outgoing nucleon the relation
\begin{equation}\label{1.91}
T = \omega + \frac{\omega^2}{2m}.
\end{equation}
Even if we use for the nucleon the nonrelativistic Schr\"{o}dinger equation where $T={\bf p}^2/2m$,
we obtain from (\ref{1.91}) the correct relativistic relation between the energy-transfer
$\omega$ and the momentum ${\bf p}$ of the nucleon in the continuum.
\begin{equation}\label{1.92}
\omega = \sqrt{m^2 +{\bf p}^2} - m.
\end{equation}
So, the main effect of this substitution is pure kinematical. There is, however
another effect. It 
is the presence of the negative energy intermediate states in the loop. However, in
our region of $\omega < 2m$ this part is real and does not contribute to the
cross-section.

\begin{figure}
\begin{center}
{ \parbox[t]{6cm}{\epsfxsize 6cm \epsffile{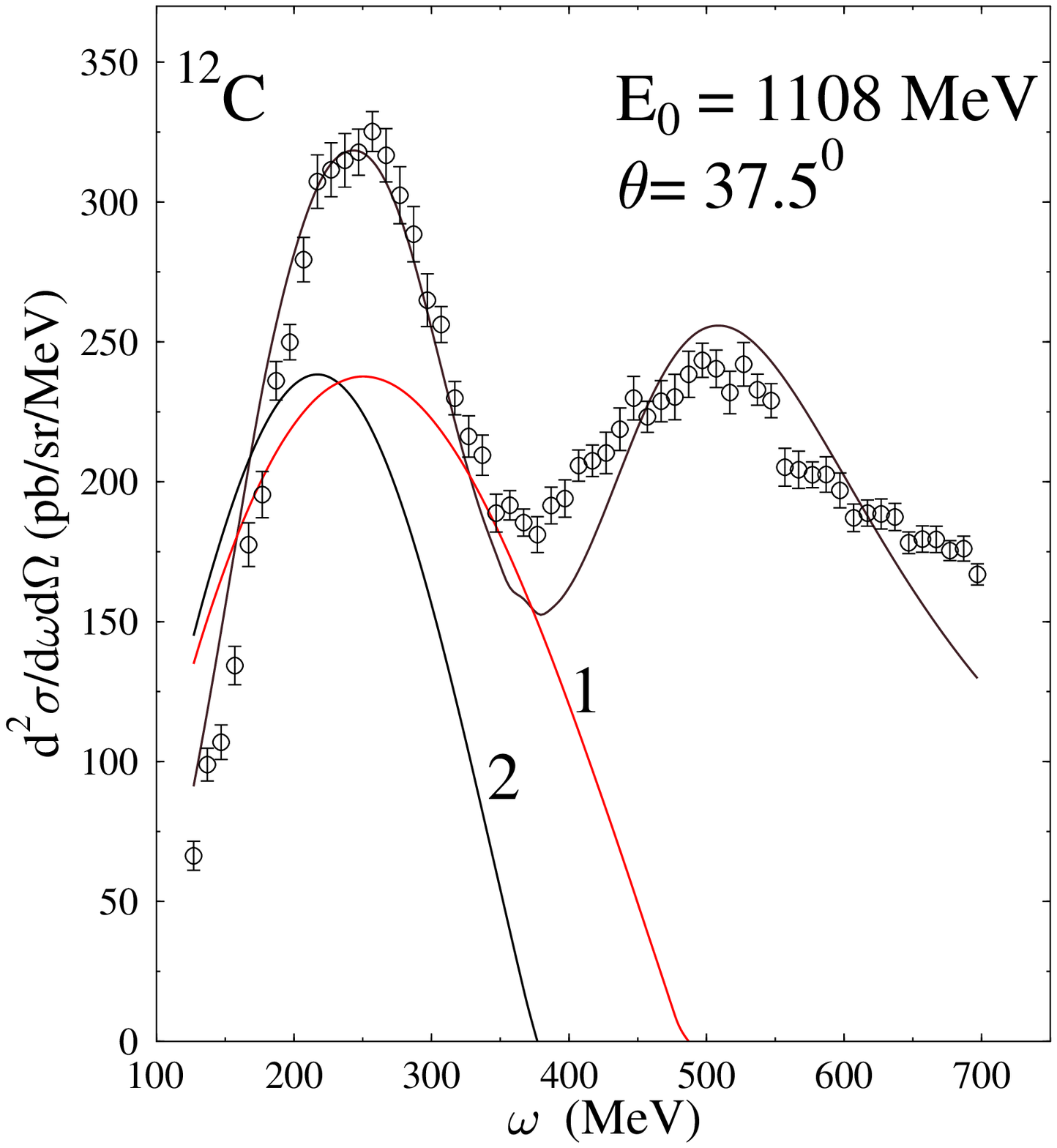}
\caption{1 - nonrelativistic Fermi-gas, 2 - relativistic Fermi-gas}
}
\hspace{3mm} \parbox[t]{6cm}{\epsfxsize 6cm \epsffile{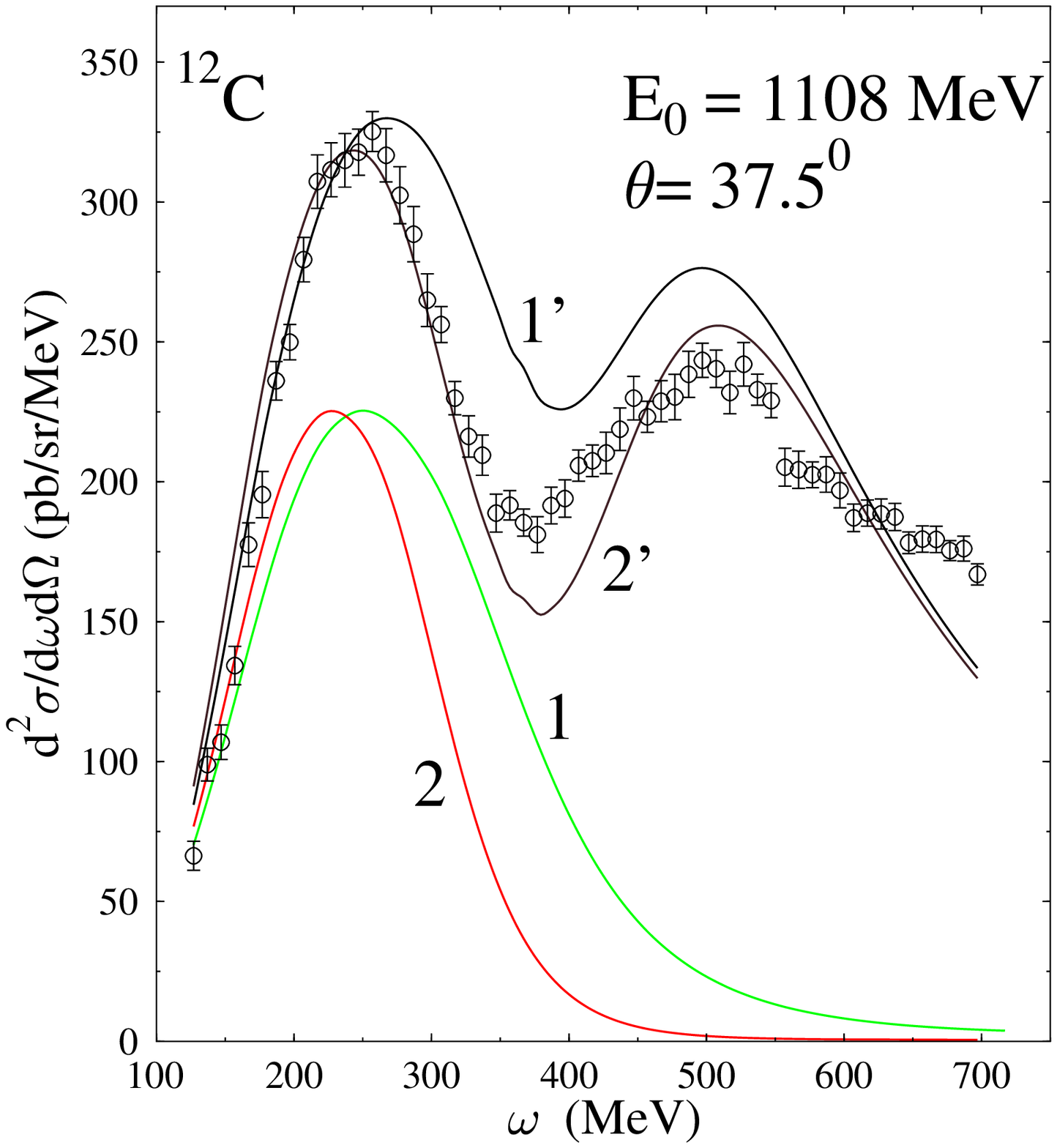}
\caption{1, 1' - finite size nucleus, 2, 2' - "quasi-relativistic" finite size nucleus}
} }
\end{center}
\end{figure}

Since the main effect has pure kinematical origin, we can try to extend our calculations for the
finite size response functions to the region of high momentum transfer just by substituting
$\omega$ in the argument of the Green functions in (\ref{1.6}) by $\omega + \omega^2/2m$. This
substitution will improve the energy-momentum relation for outgoing nucleons stabilizing, thus,
the quasi-elastic peak. In Fig.3 we show the results of our calculations of the response functions
 for the finite size nucleus where we, first, substituted $\omega \rightarrow \omega + \omega^2/2m$,
second, redefined the formfactors $F_e^2 \rightarrow A({\bf q}^2,\omega)$ for the Coulomb response
and changed
\begin{equation}\label{1.21}
F_e^2 \to \frac{F_e^2+\eta F_m^2}{1+ \eta}, \;\;  F_m^2 \to
F_m^2 \Big [\frac{-q^2}{{\bf q}^2} \Big ],
\end{equation}
for the transversal response function. These last changes do not influence much the result, except
for the Coulomb response of neutrons. The main effect comes from the improved kinematics.
With these changes the result of finite size calculations does not differ much from the 
relativistic Fermi gas, except the low-energy wing of the quasi-elastic peak as it was found
in \cite{amaro96}. One should keep in mind that the change (\ref{1.21}) in the formfactors
can be justified for the spin saturated nuclei only. For other nuclei one has to use the
explicit relativistic corrections for the electromagnetic currents.

For a finite nucleus the response functions were calculated in the independent particles model
where the particles were moving in the Woods-Saxon potential \cite{BM}
\begin{equation}\label{1.22}
U(r)=(V +\imath W)f(r) + V_{ls}r_0^2({\bf ls})\frac{1}{r}\frac{df(r)}{dr} + V_C(r),
\end{equation}
where
$$
f(r)=\frac{1}{1 + \exp{(r-R)/a}}.
$$

The inclusion of the meson-exchange currents and the final state interaction for the outgoing
nucleons leads to inevitable use of the optical potential for nucleons in the continuum 
\cite{hori80}.
In the region of the quasi-elastic peak we found the value of the imaginary part of the
optical potential (\ref{1.22}) $W = 13$ MeV. This value is consistent with the parameters of the
optical potential found in elastic proton scattering at the same proton energy \cite{fe69}.

The imaginary part of the optical potential brings an additional problem. It violates the 
conservation of the electromagnetic current. This nonconservation is very natural, since the
imaginary part of the optical potential describes the flow of nucleons from single-particle to
many-particle configurations. However, the amplitude of the process remains gauge invariant. 
In the amplitude the hadronic current is multiplied by the conserving electron electromagnetic
current and the gauge dependent part of the amplitude, which is proportional to $q_\mu$,
disappears. The Coulomb response function $S_c(w,{\bf q})$ depends explicitly now on both
the charge density and the space-longitudinal part of the current density. However, 
we can redefine the hadronic current subtracting its 4-dimensionally longitudinal part
\begin{equation}\label{1.24}
J^\prime_\mu=J_\mu-2\frac{q_\mu}{q^2}\cdot \Psi ^\dagger ({\bf r})
W({\bf r}) \Psi({\bf r})         
\end{equation} 
For this corrected current we have the same expressions for the $S_C(\omega,{\bf q})$ via 
$J^\prime_\mu$ as given by (\ref{1.2}).

\section{$\Delta$-isobar excitation}

       The $\Delta$ - production contributes to the $S_T(\omega,{\bf q})$.
The elementary amplitude of the $\Delta$ - production is given by [8]
\begin{equation}\label{1.25}
T_ {\gamma N,\Delta}(q^2)=F_N(q^2) \cdot (1+q^2/t^2)
\frac{f_{\gamma N,\Delta}}{m_ \Delta}
\sum \limits _j i[{\bf S}_j \times{\bf q}]T_{3j}e^{i{\bf qr}_j},
\end{equation}
where $F_N(q^2)$ is the usual nucleon electromagnetic formfactor, and
$t=6 GeV/c$. The coupling constant $f_{\gamma N,\Delta}$=3.6 was found 
from the reaction on a single proton \cite{ang96}.
${\bf S}$ and ${\bf T}$ are the spin and the isospin transition matrices.

        For a spherical nucleus it is convenient to use the multipole
expansion
\begin{equation}\label{1.26}
[{\bf S}_j \times {\bf q}] _ \nu e^{\imath{\bf qr}}= 4 \pi
\sum \limits_{JLM}(i)^L \j _L(qr)({\bf S} \cdot
{\bf Y}^L_{JM}(\hat{\bf n}))[{\bf Y}_{JM}^{L \ast} (\hat{\bf q}) \times {\bf q}]_ \nu
\end{equation}
where ${\bf Y}_{JM}^L({\bf n})$ is the vector spherical function. The multipole
transition density in terms of the tensor operators (\ref{1.26}) is defined by
\begin{equation}\label{1.27}
\rho^L_{JM}(r)= \sum \limits_j({\bf S} \cdot {\bf Y}_{JM}^L({\bf n}))T_{3j}
\frac{\delta(r-r_j)}{r^2} 
\end{equation}

        The structure function $S_T(\omega,{\bf q})$ can be presented then as a
sum of contributions from different multipoles
$$
S_T(\omega,{\bf q})=F^2_N(q^2)(1+q^2/t^2)^2
\frac{4 \pi F^2_ {\gamma N \Delta}(q^2)}{m^2_\Delta}
\sum \limits_{J=0}^ \infty [(J+1)W^J_{J-1J-1}(\omega,q)+
$$
$$
JW^J_{J+1J+1}(\omega, q)- \sqrt{J(J+1)}W_{J-1J+1}^J(\omega,q)-
$$
\begin{equation}\label{1.28}
\sqrt{J(J+1)}W^J_{J+1J-1}(\omega, q)+(2J+1)W^J_{JJ}(\omega, q)],
\end{equation}
where $W^J_{LL'}(\omega, q)$ is the diagonal Fourie component of the
imaginary part of the response function
\begin{equation}\label{1.29}
W^J_{LL'}(\omega, q)=- \frac{1}{\pi}
\int _0^ \infty r^2drj_L(qr)ImR_{LL'}^J(r,r';
\omega)j_{L'}(qr')r^{'2}dr',
\end{equation}

\begin{equation}\label{1.30}
\delta_{JJ'} \delta_{MM'}R^J_{LL'}(r,r'; \omega)=
<i| \rho^{\dag L}_{JM}(r)
\frac{1}{\omega -E_i-H+i \delta} \rho^{L'}_{J'M'}(r')|i>.
\end{equation}
The many-body hamiltonian $H$ in (\ref{1.30}) includes the $\Delta$ - nucleus and nucleon - nucleus single particle hamiltonians together with
free $\Delta$ width and the $\Delta - N$ and $N-N$ residual
interactions.
\begin{equation}\label{1.31}
H=h_ \Delta-i \Gamma_ \Delta/2+
\sum \limits ^{A-1}_{j=1}
\sum \limits^{A-1}_{j=1}V_{\Delta N}(r_ \Delta -r_j)+ \frac{1}{2}
\sum \limits^{A-1}_{i \ne j}V_{NN}(r_i-r_j)
\end{equation}

        The transition density operator (\ref{1.27}) acting on ground nuclear
state creates a $\Delta$ - hole state. Using this set of states and
neglecting the residual interaction one can obtain the following
expression for the response function (\ref{1.30})
$$
R^J_{LL'}(r,r'; \omega)=
\frac{2}{3} \frac{1}{2J+1} \times
$$
\begin{equation}\label{1.32}
\sum \limits _{\nu_ \Delta \nu_h}n_{\nu_ h}
\frac{< \nu_h \|{\bf S}^\dag \cdot {\bf Y}^{L*}_J \| \nu_ \Delta >
<\nu_ \Delta \|{\bf S} \cdot {\bf Y}_J^L \| \nu_ h>
R^*_{\nu_ h}(r) R_{\nu_ \Delta}(r)
R^*_{\nu_ \Delta} (r') R_{\nu_h}(r')}
{\omega+ \epsilon_{\nu_ h}- \epsilon_{\nu_ \Delta}-
\Delta m+i \Gamma_ \Delta (\omega)/2}
\end{equation}
where $\epsilon_ \nu$ and $R_ \nu(r)$ are the single-particle
energies and the wave functions of the $\Delta$ and the bounded nucleons.
In (\ref{1.32}) we neglected the contribution of the backward loop because it has
large energy denominator, of the order of $2 \Delta m$, and it
does not have the imaginary part at all.

\begin{figure}
\epsfxsize 12cm \epsfysize 8cm \epsffile{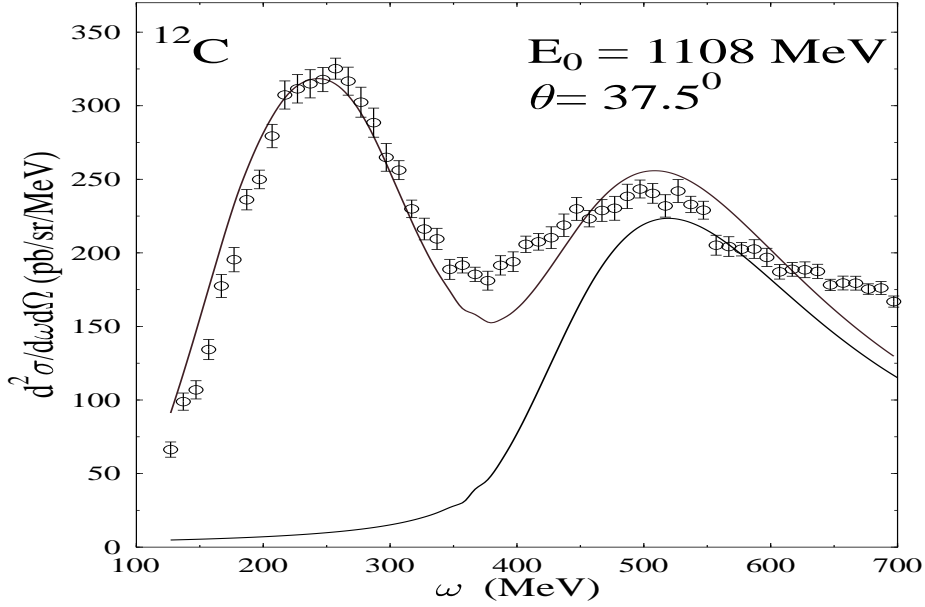}
\caption{The $\Delta$-isobar contribution}
\end{figure} 

        In order to calculate the response function (\ref{1.32}) of a finite
nucleus it is convenient to define the Green function of the
single-particle radial equation for a $\Delta$
\begin{equation}\label{1.33}
G_{J_ \Delta l_ \Delta}(r,r'; \epsilon)=
\sum \limits_{n_ \Delta}
\frac{R_{n_ \Delta j_ \Delta l_ \Delta}(r)
R^*_{n_ \Delta j_ \Delta l_ \Delta}(r')}
{\epsilon- \epsilon_{n_ \Delta j_ \Delta l_ \Delta}+
i \Gamma(\epsilon)/2}
\end{equation}
that satisfies the equation
\begin{equation}\label{1.34}
(\epsilon- \stackrel{\land}{h}_{j_ \Delta l_ \Delta})
G_{j_ \Delta l_ \Delta}(r,r'; \epsilon)= \delta(r-r'),
\end{equation}
where $\stackrel{\land}{h}_{j_ \Delta l_ \Delta}$ is the radial
$\Delta$ - nucleus Hamiltonian. The asymptotic behavior of the
$G_{j_ \Delta l_ \Delta}(r,r'; \epsilon)$ at large $r,r'$ is
determined by a pole position in (\ref{1.33}).
\begin{equation}\label{1.35}
G_{j_ \Delta l_ \Delta}(r,r') \sim \exp (ikr),
\end{equation}
where $k= \sqrt{2m_ \Delta / \hbar^2(\epsilon+i \Gamma /2)}$. The
expression for the response function (\ref{1.32}) becomes
$$
R^J_{LL'}(r,r'; \omega)=
\frac{1}{2}\frac{1}{2J+1}
\sum \limits_{\nu_ \Delta \nu_ h}n_{\nu_ h}
<\nu_ h \|{\bf S}^\dag \cdot{\bf Y}^{L*}_J \| \nu_ \Delta>
<\nu_ \Delta \|{\bf S} \cdot{\bf Y}^L_J \| \nu_ h> \times
$$
\begin{equation}\label{1.36}
G_{j_ \Delta l_ \Delta}(r,r'; \omega- \Delta m+ \epsilon_{\nu_ h}).
\end{equation}

The $\Delta$-nucleus optical potential has been taken in the Woods-Saxon form with the
parameters found from the pion-nucleus data \cite{horik80}
The results of the calculations of the $\Delta$-isobar contribution into inclusive
spectrum for the electron scattering from $^{12}C$ at the energy of incident electron
1108 MeV are shown in the Fig.4.

\section{Meson-exchange currents}

It is known that the single-nucleon and single-$\Delta$ degrees of 
freedom do not exhaust the spectrum especially in the intermediate 
region between N and $\Delta$ peaks. The meson-exchange currents are
necessary to account for quantitative description of the spectra
\cite{ris89,vod81}. Part of the MEC has been included already into the 
$\Delta$-excitation process. Here we shall discuss only the nucleon
sector of MEC. 
 
        As it can be inferred from \cite{ris89} there are three types of
diagrams shown in Fig.5. This diagrams correspond to the following expressions for
the current densities: 
$$        
J_\mu^{ \pi -in-fl}=2iP_\mu \;
({\bf P}-{\bf q}/2)\cdot\mbox{\boldmath $\sigma$}_{11'}\;
({\bf P}+{\bf q}/2) \cdot\mbox{\boldmath $\sigma$}_{22'}
\cdot 
$$
\begin{equation}\label{1.37}
 [\mbox{\boldmath $\tau$}_1 \times \mbox{\boldmath $\tau$}_2]_3 \cdot
\frac{f_\pi(P-q/2)  \cdot f_\pi(P+q/2)}{({m_\pi}^2-(P+q/2)^2) 
({m_\pi}^2-(P-q/2)^2)},
\end{equation} 
Fig.5 (c), which is usually called the pion-in-flight contribution.
$$        
J_\mu^{cont}=i[\overline{U}(P_2') 
(\hat{P}+\hat{q}/2) \gamma_5 U(P_2) \,
\overline{U}(P_1') \gamma_\mu \gamma_5 U(P_1) \,
[\mbox{\boldmath $\tau$}_1 \times \mbox{\boldmath $\tau$}_2]_3
\cdot
$$
$$
 \frac{{f_\pi}^2(P+q/2)}{({m_\pi}^2-(P+q/2)^2)} + 
\overline{U}(P_1') 
(\hat{P}-\hat{q}/2) \gamma_5 U(P_1) \,
\overline{U}(P_2') \gamma_\mu \gamma_5 U(P_2) \,
[\mbox{\boldmath $\tau$}_1 \times \mbox{\boldmath $\tau$}_2]_3
$$
\begin{equation}\label{1.38}
\cdot \frac{{f_\pi}^2(P-q/2)}
{({m_\pi}^2-(P-q/2)^2)}] 
\end{equation}

\begin{figure}
\begin{center}
{ \parbox[t] {3cm}{ \begin{picture}(80,90)
            \multiput(20,75)(-10,10){3}{\oval(10,10)[bl]}
            \multiput(10,75)(-10,10){3}{\oval(10,10)[tr]}
            \thicklines
            \multiput(0,20)(0,50){2}{\vector(1,0){30}}
            \multiput(50,20)(0,50){2}{\vector(1,0){30}}
            \multiput(30,20)(0,50){2}{\line(1,0){20}}
            \multiput(40,20)(0,20){3}{\line(0,1){10}}
            \put(60,10){${p_2}'$}
            \put(5,10){${p_2}$}
            \put(5,60){${p_1}$}
            \put(60,60){${p_1}'$}
            \put(10,85){$q$}
            \put(30,40){$p$}
            \put(165,-5){(a)}
            \end{picture}}
\parbox[t] {3cm} {  \begin{picture}(80,90)
            \multiput(55,75)(-10,10){3}{\oval(10,10)[bl]}
            \multiput(45,75)(-10,10){3}{\oval(10,10)[tr]}
            \thicklines
            \multiput(0,20)(0,50){2}{\vector(1,0){30}}
            \multiput(50,20)(0,50){2}{\vector(1,0){30}}
            \multiput(30,20)(0,50){2}{\line(1,0){20}}
            \multiput(40,20)(0,20){3}{\line(0,1){10}}
            \put(60,10){${p_2}'$}
            \put(5,10){${p_2}$}
            \put(5,60){${p_1}$}
            \put(60,60){${p_1}'$}
            \put(50,85){$q$}
            \put(30,40){$p$}
            \end{picture}}
\parbox[t] {3cm} {\begin{picture}(80,90)
            \multiput(55,75)(-10,10){3}{\oval(10,10)[bl]}
            \multiput(45,75)(-10,10){3}{\oval(10,10)[tr]}
            \thicklines
            \multiput(0,20)(0,50){2}{\vector(1,0){30}}
            \multiput(50,20)(0,50){2}{\vector(1,0){30}}
            \multiput(30,20)(0,50){2}{\line(1,0){20}}
            \multiput(40,20)(0,20){3}{\line(0,1){10}}
            \put(60,10){${p_1}'$}
            \put(5,10){${p_1}$}
            \put(5,60){${p_2}$}
            \put(60,60){${p_2}'$}
            \put(50,85){$q$}
            \put(30,40){$p$}
            \end{picture}}
\parbox[t] {3cm} {\begin{picture}(80,90)
            \multiput(55,75)(-10,10){3}{\oval(10,10)[bl]}
            \multiput(45,75)(-10,10){3}{\oval(10,10)[tr]}
            \thicklines
            \multiput(0,20)(0,50){2}{\vector(1,0){30}}
            \multiput(50,20)(0,50){2}{\vector(1,0){30}}
            \multiput(30,20)(0,50){2}{\line(1,0){20}}
            \multiput(40,20)(0,20){3}{\line(0,1){10}}
            \put(60,10){${p_1}'$}
            \put(5,10){${p_1}$}
            \put(5,60){${p_2}$}
            \put(60,60){${p_2}'$}
            \put(50,85){$q$}
            \put(30,40){$p$}
            \end{picture}}
\vspace{4mm}
\parbox[t] {4cm}{\begin{picture}(80,90)
            \multiput(43,75)(-10,10){3}{\oval(10,10)[bl]}
            \multiput(33,75)(-10,10){3}{\oval(10,10)[tr]}
            \thicklines
            \multiput(0,20)(0,50){2}{\vector(1,0){30}}
            \multiput(50,20)(0,50){2}{\vector(1,0){30}}
            \multiput(30,20)(0,50){2}{\line(1,0){20}}
            \multiput(40,20)(0,20){3}{\line(0,1){10}}
            \put(60,10){${p_2}'$}
            \put(5,10){${p_2}$}
            \put(5,60){${p_1}$}
            \put(60,60){${p_1}'$}
            \put(35,85){$q$}
            \put(30,40){$p$}
            \put(90,-5){(b)}
            \end{picture}}  
\parbox[t] {4cm} {\begin{picture}(80,90)
            \multiput(43,75)(-10,10){3}{\oval(10,10)[bl]}
            \multiput(33,75)(-10,10){3}{\oval(10,10)[tr]}
            \thicklines
            \multiput(0,20)(0,50){2}{\vector(1,0){30}}
            \multiput(50,20)(0,50){2}{\vector(1,0){30}}
            \multiput(30,20)(0,50){2}{\line(1,0){20}}
            \multiput(40,20)(0,20){3}{\line(0,1){10}}
            \put(60,10){${p_1}'$}
            \put(5,10){${p_1}$}
            \put(5,60){${p_2}$}
            \put(60,60){${p_2}'$}
            \put(35,85){$q$}
            \put(30,40){$p$}
            \end{picture}}
\parbox[t] {4cm} {\begin{picture}(80,90)
            \multiput(12,43)(10,0){3}{\oval(5,5)[b]}
            \multiput(17,43)(10,0){3}{\oval(5,5)[t]}
            \thicklines
            \multiput(0,20)(0,50){2}{\vector(1,0){30}}
            \multiput(50,20)(0,50){2}{\vector(1,0){30}}
            \multiput(30,20)(0,50){2}{\line(1,0){20}}
            \multiput(40,20)(0,20){3}{\line(0,1){10}}
            \put(60,10){${p_2}'$}
            \put(5,10){${p_2}$}
            \put(5,60){${p_1}$}
            \put(60,60){${p_1}'$}
            \put(27,50){$q$}
            \put(45,30){$p$}
            \put(30,-5){(c)}
        \end{picture}}
}
\caption{Meson-exchange currents diagrams}
\end{center}
\end{figure}
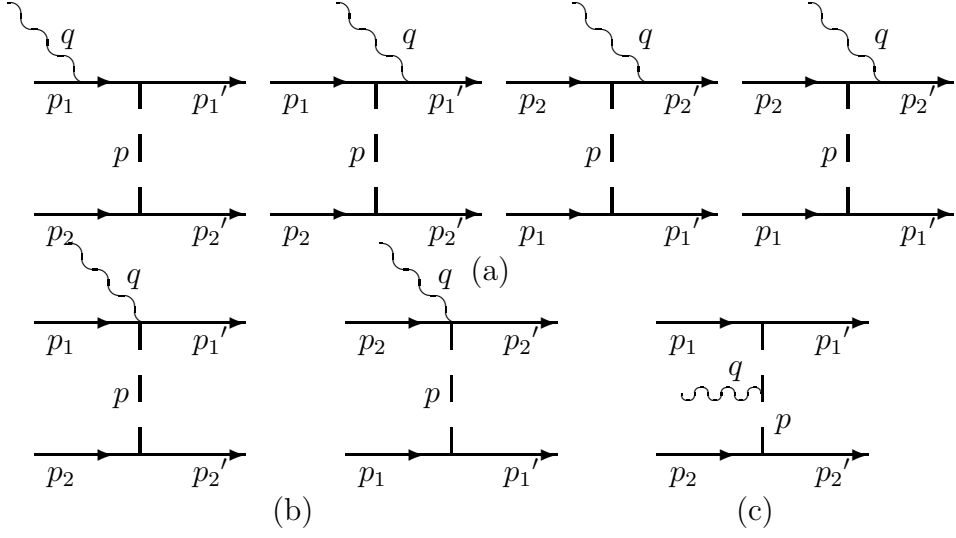

Fig.5 (b), which is called the contact contribution, and
$$
J_\mu^{prop} =[\overline{U}(P_1') \gamma _\mu
\frac{1}{(m- (\hat{P}_1'-\hat{q}))} \gamma_\nu \gamma_5 U(P_1) \cdot
\overline{U}(P_2') \gamma_\kappa \gamma_5 U(P_2) \cdot
$$
$$
\cdot (P+q/2)^\nu (P+q/2)^\kappa \cdot 
\frac{1}{2}(1+ \tau_3^{(1)}) \mbox{\boldmath $\tau$}^{(1)} \cdot
\mbox{\boldmath $\tau$}^{(2)}+
$$
$$
+ \overline{U}(P_1') \gamma_\nu \gamma_5
\frac{1}{(m-(\hat{P}_1+\hat{q}))}
\gamma_ \mu U(P_1) \cdot 
\overline {U}(P_2') \gamma_ \kappa \gamma _5
U(P_2) \cdot
$$
$$
(P+q/2)^\nu (P+q/2)^\kappa \cdot\mbox{\boldmath $\tau$}^{(1)} \cdot
\mbox{\boldmath $\tau$}^{(2)}
 \frac{1}{2}(1+ \tau_3^{(1)})] \cdot
\frac{{f_\pi}^2(P+q/2)}{({m_\pi}^2-(P+q/2)^2)} +
$$
\begin{equation}\label{1.39}
+ [1 \leftrightarrow 2, q \leftrightarrow -q ] \cdot 
\frac{{f_\pi}^2(P-q/2)}{({m_\pi}^2-(P-q/2)^2)}
\end{equation} 
Fig.5 (a), which is the propagation contribution.
The factor $f_\pi(P)$ is a pion formfactor and it was taken as in
\cite{ris89}
$$
f_\pi(p)=(\Lambda^2-m_\pi^2)/(\Lambda^2-p^2)
\eqno(36)
$$.

In our calculations the best agreement with the data was obtained for
 $\Lambda$ = 1250 MeV. 

        The total current, which is the sum of these three terms $
J_ \mu = J_ \mu^{\pi-in-fl}+J_ \mu^{cont}+
J_ \mu^{prop} $ , is really conserving. 

Let us discuss the diagram (a) more closely. Here we have two different
situations for the positive and the negative energy states in the propagator 
of nucleons.
We would like to note, that the diagrams have a pole in the region
of the quasi-elastic peak and we inevitably cross it during the integration
over the initial momentum of the nucleon. This singularity produces infinite 
contribution to the cross-section. The way of treating this singularity in given
order of perturbation theory has been indicated in \cite{alb90}. We would like to 
use different approach in the interpretation of the pole. Let us remark
that the pion-exchange is not specific in this diagram and it can be replaced by any other 
interaction. The pole in the diagrams does not disappear. One can immediately draw the higher
order diagrams, where the singularity will be accumulating. As it was shown
in \cite{hori80}, this kind of diagrams corresponds to the interaction in 
final state for the positive energy states of propagating nucleon. Summing 
these diagrams we come to the optical
 potential  for the final state nucleons and the contribution of the 
 diagrams (a) for the 
intermediate states with  positive energy should be omitted at all if we use an optical
potential in order to avoid double counting.

As for negative energy intermediate states, it was shown in \cite{vod81} that
their contribution is small and we neglect it here as well. 

In order to calculate the cross-section we have to square the obtained currents 
and to integrate them over three momenta $P$, $P_1$, and $P_2$. The simplest way 
to do it is to calculate the corresponding response functions as it has been done 
for the single-nucleon channel. The diagrams for the response functions are
presented in Fig.6. It is seen clearly that the integration over the momenta is 
almost factorized. We have separate integration in the upper loop over ($P_1$),
in the lower loop over ($P_2$), and over the momentum carried by the pion ($P$).

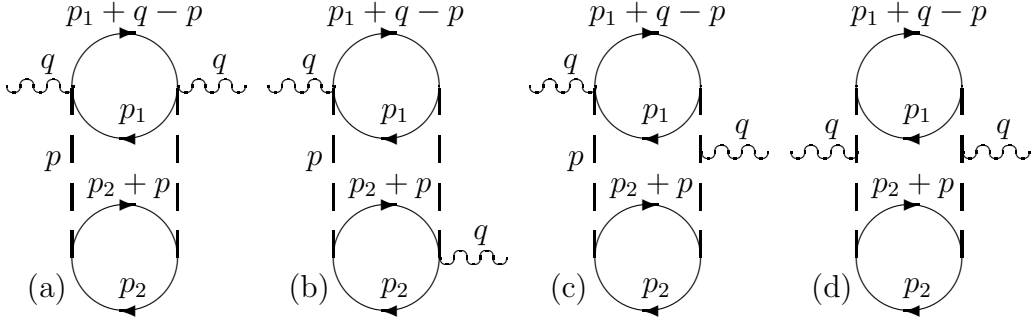
\begin{figure}
\begin{center}
{ \parbox[t] {3cm} {\begin{picture}(80,90)
            \multiput(0,75)(10,0){3}{\oval(5,5)[b]}
             \multiput(5,75)(10,0){2}{\oval(5,5)[t]}
            \multiput(65,75)(10,0){3}{\oval(5,5)[b]}
            \multiput(70,75)(10,0){2}{\oval(5,5)[t]}
            \multiput(42,75)(-10,10){1}{\oval(40,40)[b]}
            \multiput(42,75)(-10,10){1}{\oval(40,40)[t]}
            \multiput(42,10)(-10,10){1}{\oval(40,40)[b]}
            \multiput(42,10)(-10,10){1}{\oval(40,40)[t]}
      \thicklines
            \multiput(43,95)(0,50){1}{\vector(1,0){3}}
            \multiput(43,55)(0,50){1}{\vector(-1,0){3}}
           \multiput(43,30)(0,50){1}{\vector(1,0){3}}
            \multiput(43,-10)(0,50){1}{\vector(-1,0){3}}
           \multiput(22,10)(0,18){4}{\line(0,1){10}}
            \multiput(62,10)(0,18){4}{\line(0,1){10}}
            \put(28,35){${p_2+p}$}
            \put(40,-3){${p_2}$}
            \put(40,63){${p_1}$}
            \put(20,100){${p_1+q-p}$}
            \put(10,82){$q$}
            \put(12,45){$p$}
           \put(75,82){$q$}
           \put(5,-5){(a)}
            \end{picture}}
\hspace{2mm} \parbox[t] {3cm}{\begin{picture}(80,90)
            \multiput(0,75)(10,0){3}{\oval(5,5)[b]}
             \multiput(5,75)(10,0){2}{\oval(5,5)[t]}
            \multiput(65,10)(10,0){3}{\oval(5,5)[b]}
            \multiput(70,10)(10,0){2}{\oval(5,5)[t]}
            \multiput(42,75)(-10,10){1}{\oval(40,40)[b]}
            \multiput(42,75)(-10,10){1}{\oval(40,40)[t]}
            \multiput(42,10)(-10,10){1}{\oval(40,40)[b]}
            \multiput(42,10)(-10,10){1}{\oval(40,40)[t]}
      \thicklines
            \multiput(43,95)(0,50){1}{\vector(1,0){3}}
            \multiput(43,55)(0,50){1}{\vector(-1,0){3}}
           \multiput(43,30)(0,50){1}{\vector(1,0){3}}
            \multiput(43,-10)(0,50){1}{\vector(-1,0){3}}
           \multiput(22,10)(0,18){4}{\line(0,1){10}}
            \multiput(62,10)(0,18){4}{\line(0,1){10}}
            \put(28,35){${p_2+p}$}
            \put(40,-3){${p_2}$}
            \put(40,63){${p_1}$}
            \put(20,100){${p_1+q-p}$}
            \put(10,82){$q$}
            \put(12,45){$p$}
           \put(75,17){$q$}
           \put(5,-5){(b)}
            \end{picture}}
\hspace{2mm} \parbox[t] {3cm} {\begin{picture}(80,90)
            \multiput(0,75)(10,0){3}{\oval(5,5)[b]}
             \multiput(5,75)(10,0){2}{\oval(5,5)[t]}
            \multiput(65,50)(10,0){3}{\oval(5,5)[b]}
            \multiput(70,50)(10,0){2}{\oval(5,5)[t]}
            \multiput(42,75)(-10,10){1}{\oval(40,40)[b]}
            \multiput(42,75)(-10,10){1}{\oval(40,40)[t]}
            \multiput(42,10)(-10,10){1}{\oval(40,40)[b]}
            \multiput(42,10)(-10,10){1}{\oval(40,40)[t]}
      \thicklines
            \multiput(43,95)(0,50){1}{\vector(1,0){3}}
            \multiput(43,55)(0,50){1}{\vector(-1,0){3}}
           \multiput(43,30)(0,50){1}{\vector(1,0){3}}
            \multiput(43,-10)(0,50){1}{\vector(-1,0){3}}
           \multiput(22,10)(0,18){4}{\line(0,1){10}}
            \multiput(62,10)(0,18){4}{\line(0,1){10}}
            \put(28,35){${p_2+p}$}
            \put(40,-3){${p_2}$}
            \put(40,63){${p_1}$}
            \put(20,100){${p_1+q-p}$}
            \put(10,82){$q$}
            \put(12,45){$p$}
           \put(75,57){$q$}
           \put(5,-5){(c)}
            \end{picture}}
\hspace{2mm} \parbox[t] {3cm} {\begin{picture}(80,90)
            \multiput(0,50)(10,0){3}{\oval(5,5)[b]}
             \multiput(5,50)(10,0){2}{\oval(5,5)[t]}
            \multiput(65,50)(10,0){3}{\oval(5,5)[b]}
            \multiput(70,50)(10,0){2}{\oval(5,5)[t]}
            \multiput(42,75)(-10,10){1}{\oval(40,40)[b]}
            \multiput(42,75)(-10,10){1}{\oval(40,40)[t]}
            \multiput(42,10)(-10,10){1}{\oval(40,40)[b]}
            \multiput(42,10)(-10,10){1}{\oval(40,40)[t]}
      \thicklines
            \multiput(43,95)(0,50){1}{\vector(1,0){3}}
            \multiput(43,55)(0,50){1}{\vector(-1,0){3}}
           \multiput(43,30)(0,50){1}{\vector(1,0){3}}
            \multiput(43,-10)(0,50){1}{\vector(-1,0){3}}
           \multiput(22,10)(0,18){4}{\line(0,1){10}}
            \multiput(62,10)(0,18){4}{\line(0,1){10}}
            \put(28,35){${p_2+p}$}
            \put(40,-3){${p_2}$}
            \put(40,63){${p_1}$}
            \put(20,100){${p_1+q-p}$}
            \put(10,57){$q$}
           \put(75,57){$q$}
           \put(5,-5){(d)}
            \end{picture}}
}
\caption{Loops for MEC}
\end{center}
\end{figure}
      
The calculations of the particle-hole loop for the pion channel have been 
done in \cite{dmi85}. We need to compute the upper loops only. The result is:
for the diagram (a)
$$        
R^{( )}_{up}(\omega, {\bf q}, P)= 
\frac{2}{(2 \pi)^3} \int d^3P_1 \Big[
\frac{A({\bf q}, \omega, P, P_1)}
{\omega + P_0 + (q + P)^2/2m- ({\bf P_1} \cdot {\bf (q+P)})/m}- 
$$
\begin{equation}\label{1.41}
-\frac{A({\bf q}, \omega, P, -P_1)}
{\omega + P_0 - (q + P)^2/2m- ({\bf P_1} \cdot {\bf (q+P)})/m}
 \Big ],   
\end{equation} 
where        
$$
A({\bf q}, \omega, P, P_1)=4 \Big [(2m^2+P_1 \cdot (q-P)) \cdot(g_{\mu
\nu}-\frac{q_\mu q_\nu}{q^2}) - 
$$
$$
- 2(P_{1\mu} - \frac{q_{\mu}
(qP_1)}{q^2}) \cdot (P_{1\nu} - \frac{q_{\nu}(qP_1)}{q^2}) +
(P_{1\mu} - \frac{q_{\mu}(qP_1)}{q^2}) \cdot (P_{\nu} - \frac{q_{\nu}
(qP)}{q^2}) +
$$
\begin{equation}\label{1.42}
+ (P_{\mu} - \frac{q_{\mu}(qP)}{q^2}) \cdot (P_{1\nu} - \frac{q_{\nu}
(qP_1)}{q^2})
\end{equation}

        Here $A({\bf q}, \omega, P)$ is written in a symbolic form both for
the Coulomb and the transversal responses.

For (b) we obtain

$$        
R^{(b)}_{up}(\omega, {\bf q}, P)= 
\frac{2}{(2 \pi)^3} \int d^3P_1 \Big[
\frac{B({\bf q}, \omega, P, P_1)}
{\omega + P_0 + (q + P)^2/2m- ({\bf P_1} \cdot {\bf (q+P)})/m}- 
$$
\begin{equation}\label{1.43}
-\frac{B({\bf q}, \omega, P, -P_1)}
{\omega + P_0 - (q + P)^2/2m- ({\bf P_1} \cdot {\bf (q+P)})/m}
 \Big ],   
\end{equation} 
where        
$$
B({\bf q}, \omega, P, P_1)=4 \Big [ 2m^2 \cdot (P_{\mu} - \frac{q_{\mu}
(qP_)}{q^2})(2P_\nu - q_\nu) + 
$$
\begin{equation}\label{1.44}
+ (P_{1\mu} -
\frac{q_{\mu}(qP_1)}{q^2})(2P_\nu - q_\nu) \cdot ( {(q-P)}^2 + 2
\cdot P_1(q-P) ) \Big ]
\end{equation} 
        
Using the above expressions for the response functions we obtain      
$$
R_{C,T}(\omega,{\bf q}) = 4 \int \frac{d^4P}{(2\pi)^4} \Big [
2R^{( )}_{up}(\omega, {\bf q}, P) 
R^{( )}_{low}(\omega, {\bf q}, P) \cdot
\frac{{f_\pi(P)}^4}{{(P^2-{m_\pi}^2)}^2} +        
$$
$$
+ 2R^{(b)}_{up}(\omega, {\bf q}, P) 
R^{(b)}_{low}(\omega, {\bf q}, P) \cdot
\frac{{f_\pi(P)}^2{f_\pi(q-P)}^2}{(P^2-{m_\pi}^2)({(q-P)}^2-{m_\pi}^2)}+
$$
$$ 
+ 4R^{(c)}_{up}(\omega, {\bf q}, P) 
R^{(c)}_{low}(\omega, {\bf q}, P) \cdot
\frac{{f_\pi(P)}^3 f_\pi(q-P)}{{(P^2-{m_\pi}^2)}^2({(q-P)}^2-{m_\pi}^2)}+
$$
\begin{equation}\label{1.45}
+ R^{(d)}_{up}(\omega, {\bf q}, P) 
R^{(d)}_{low}(\omega, {\bf q}, P) \cdot
\frac{{f_\pi(P)}^2 {f_\pi(q-P)}^2}{{(P^2-{m_\pi}^2)}^2  
{({(q-P)}^2-{m_\pi}^2)}^2} \Big ]
\end{equation} 

\begin{figure}
\epsfxsize 12cm \epsfysize 8cm \epsffile{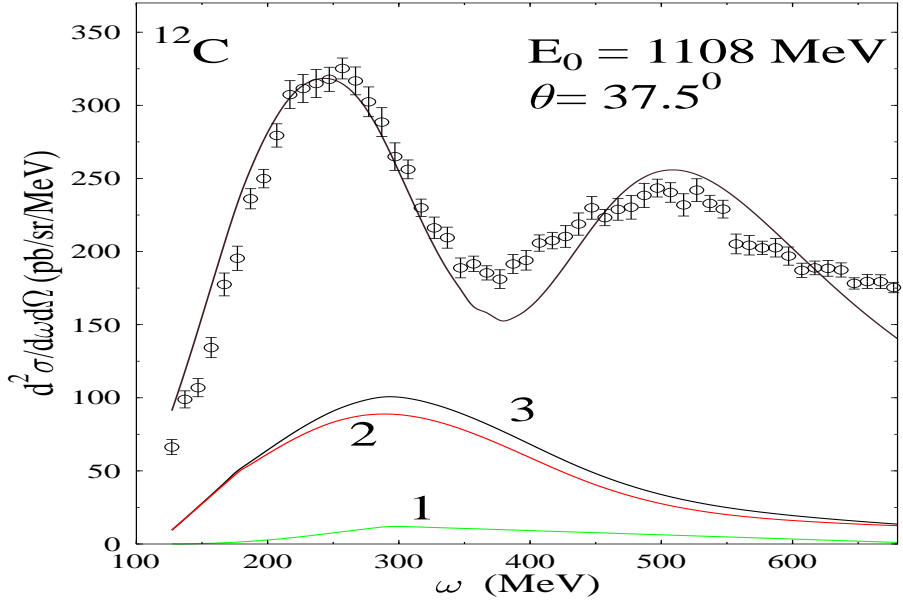}
\caption{1 - the longitudinal cross-section, 2 - the transversal cross-section, 3 - the total 
cross-section from the meson-exchange currents}
\end{figure}  
 
The contribution of each response function is shown in Fig.7. As in the previous
calculations the transversal response is considerably larger than the 
longitudinal response \cite{ryck94}. 

In general, at high momentum transfer the contribution of the meson-exchange
currents is considerable and can not be neglected.

\section{Results for $^{12}C$ and $^{16}O$}        

\begin{figure}
\begin{center}
{ \parbox[t]{5cm}{\epsfxsize 5cm \epsffile{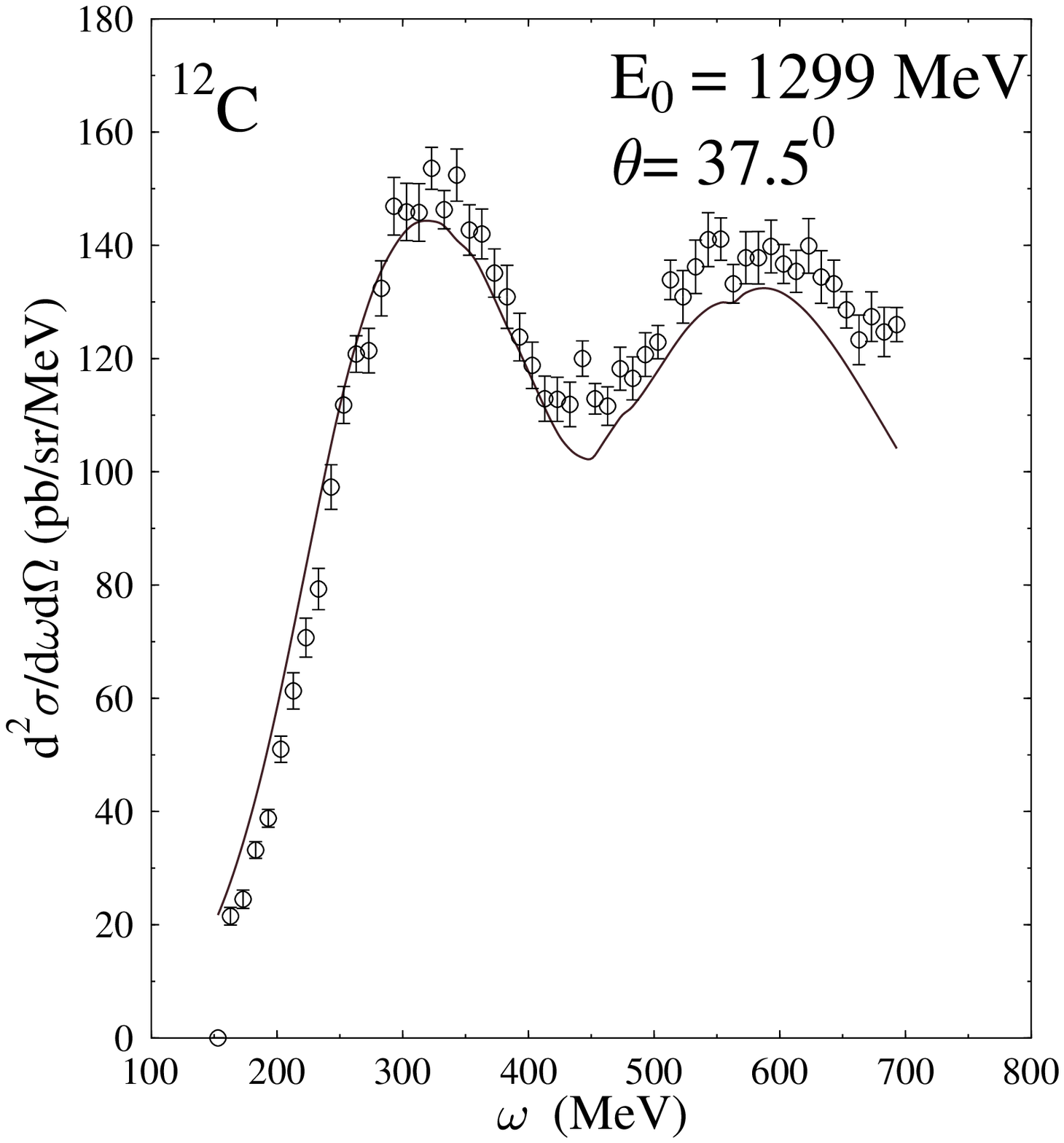}}
\hspace{3mm} \parbox[t]{5cm}{\epsfxsize 5cm \epsffile{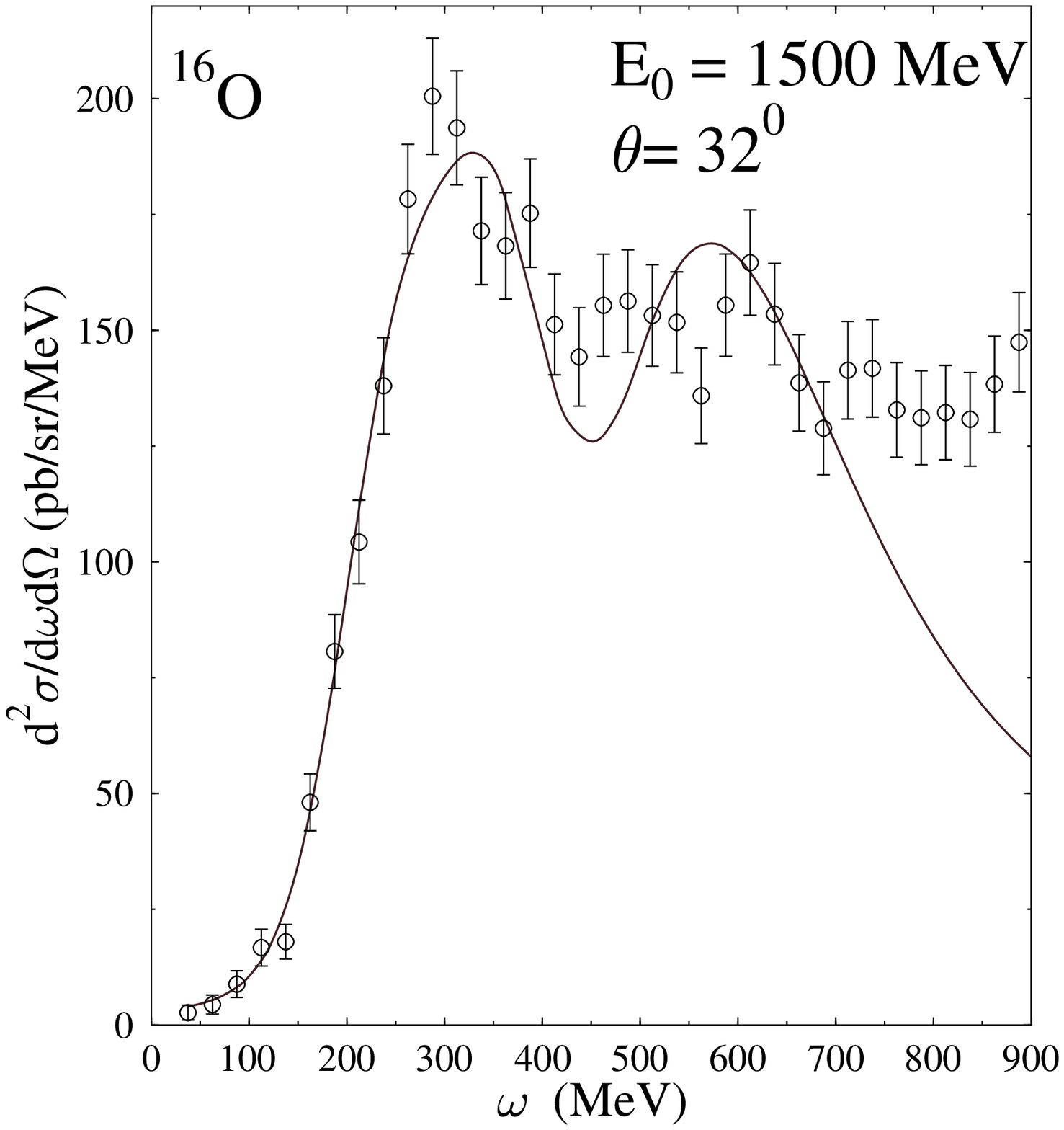}}

\parbox[t]{5cm}{\epsfxsize 5cm \epsffile{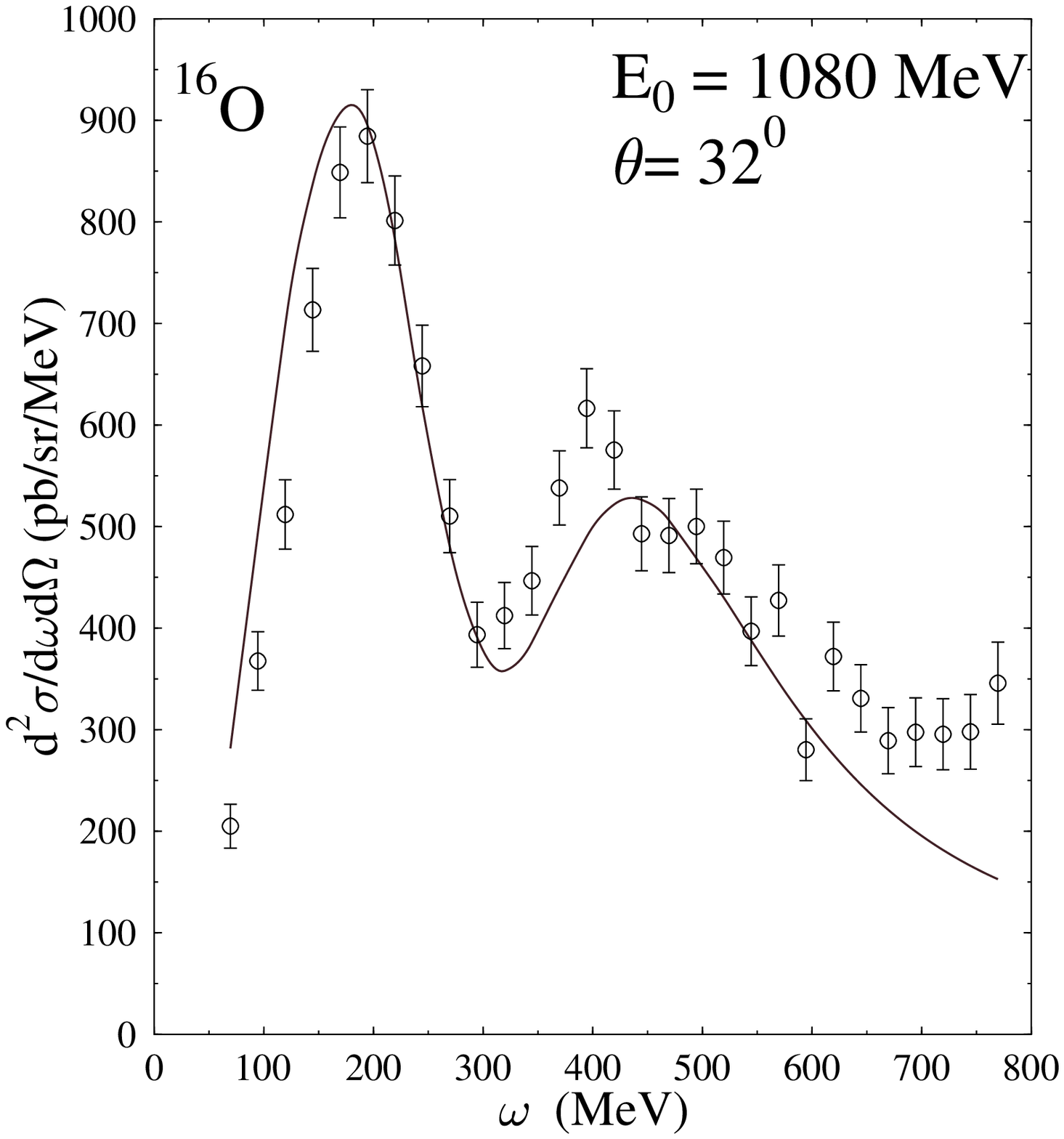}}
\hspace{3mm} \parbox[t]{5cm}{\epsfxsize 5cm \epsffile{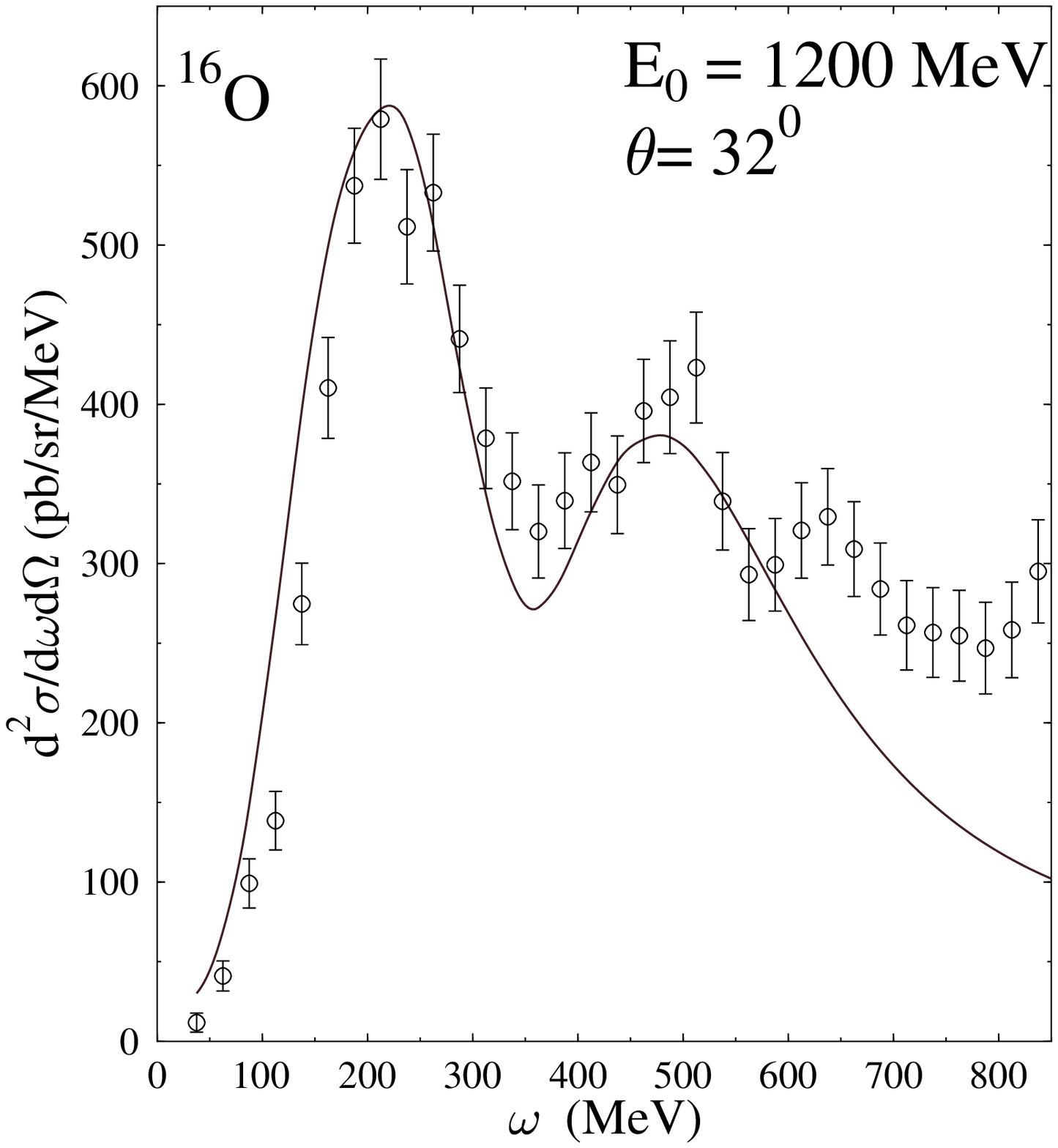}}
 }
\caption{Total cross-section including quasi-elastic peak, $\Delta$ - excitation 
and MEC} 
\end{center}
\end{figure}
     
 The final results 
of our approach are shown in Fig.8, where we present the spectra at high electron 
energy for two nuclei $^{12}C$ and $^{16}O$ \cite{ocon90,ang96}. For both nuclei the 
calculations were performed with  
the same set of the parameters (except the size of a nucleus). Again, the best fit 
was obtained to the data with the highest momentum transfer, Fig.8. At lower momentum
transfer, Fig.1 and Fig.2,  there is a small systematic shift of the quasi-elastic peak 
position. This can be indication that either some 1p-1h correlations still present in 
the final state even at this high momentum transfer, or it can be the difference in the
potential depths for the bounded nucleons and the nucleons in the continuum. The oxigen 
data show the large contribution to the cross-section behind the $\Delta$-peak. This is
the indication of higher nucleon resonances excitation. However, just the number of MEC 
diagrams is much larger in this region and they can hide the resonance contribution.
This problem needs separate investigation. 

In conclusion, we calculated the inclusive cross-section in the regions of 
quasi-elastic peak, $\Delta$ - peak and the intermediate region. It was demonstrated 
that in accordance with the earlier calculations the nonrelativistic approach fails to
reproduce the shape of quasi-elastic peak at high momentum transfer
 and the simple recipe was proposed  to account the
relativistic kinematics for the finite nucleus calculations.  

\section{Acknowledgment}

The encouraging discussions with T.-S.H. Lee on the initial stage of work are greatly
acknowledged by one of the author (V.D.) The authors are grateful to Dr. M.Anghinolfi for 
presenting the oxygen data.

\end{document}